\documentclass[aps,twocolumn,superscriptaddress,floatfix,nofootinbib]{revtex4-1}
\usepackage{graphicx,amsmath,amssymb,verbatim,color}
\usepackage{booktabs}
\usepackage[dvipsnames]{xcolor}
\usepackage{bm}
\usepackage[utf8]{inputenc}
\usepackage[colorlinks=true,citecolor=blue,linkcolor=blue,urlcolor=blue, backref=false,pdfborder={0 0 0}]{hyperref}
\usepackage{float}
\usepackage{multirow}
\usepackage{stackengine}
\usepackage{amssymb}
\usepackage{placeins}
\usepackage{afterpage}

\newcommand{\be}{\begin{equation}\begin{gathered}}
\newcommand{\ee}{\end{gathered}\end{equation}}
\newcommand{\overliner}{\begin{eqnarray}}
\newcommand{\earr}{\end{eqnarray}}

\newcommand{\ols}[1]{\mskip.5\thinmuskip\overline{\mskip-.5\thinmuskip {#1} \mskip-.5\thinmuskip}\mskip.5\thinmuskip} 

\begin{document}

\pagenumbering{arabic}

\title{Compressed \emph{Gaussian} likelihood for the \textit{Planck} low-$\ell$ data}

\author{Nanoom Lee}
\email{nanoom.lee@jhu.edu}
\affiliation{William H. Miller III Department of Physics \& Astronomy, Johns Hopkins University, Baltimore, Maryland 21218, USA}

\date{\today}
\begin{abstract}
We present a compressed \emph{Gaussian} likelihood for the \textit{Planck} CMB low-$\ell$ E-mode polarization data, constructed from the \texttt{SRoll2} likelihood which provides a state-of-the-art constraint on the reionization optical depth $\tau$ to date. The non-Gaussian form of CMB low-$\ell$ temperature (TT) and E-mode polarization (EE) likelihoods makes them incompatible with Fisher matrix analyses that require an analytic Gaussian $\chi^2$, such as the Fisher-bias formalism and Fisher forecasts. We show that the $\chi^2$ of an offset log-normal likelihood takes a Gaussian form in the log-transformed power spectrum amplitudes, and can therefore be used directly in Fisher matrix analyses without any explicit change of variables. Building on this, we compress the \texttt{SRoll2} likelihood into a small number of piecewise offset log-normal functions and validate it against the full \texttt{SRoll2} likelihood via MCMC combined with \textit{Planck} and ACT DR6 data, finding excellent agreement across all $\Lambda$CDM parameters and in extended cosmological models. We further demonstrate that Fisher matrix uncertainty estimates from our compressed likelihood agree well with the full MCMC posteriors. We release our compressed likelihood \texttt{planck-gaussian-lowl}, a lightweight Python package incorporating the compressed low-$\ell$ TT likelihood from previous work, allowing a straightforward incorporation of the Planck CMB low-$\ell$ data into any Gaussian-likelihood-based analysis. The package is publicly available at \href{https://github.com/nanoomlee/planck-gaussian-lowl}{github.com/nanoomlee/planck-gaussian-lowl}.
\end{abstract}
\maketitle

\section{Introduction}
\label{sec:intro}

The optical depth to reionization, $\tau$, is one of the six standard parameters of the flat $\Lambda$CDM model. It is primarily constrained by Cosmic Microwave Background (CMB) E-mode polarization power spectrum on the large angular scales, which carries the imprint of Thomson scattering off free electrons during the epoch of reionization. An accurate and convenient likelihood for these data is therefore essential.

The \textit{Planck} 2018 legacy release \cite{Planck:2018nkj,Planck2018,Planck:2019nip} included the \texttt{SimAll} likelihood for low-$\ell$ EE polarization, constructed from simulations of the \textit{Planck} High Frequency Instrument (HFI) maps \cite{Planck:2018lkk}. However, the 2018 maps are known to be affected by residual instrumental systematic effects, most notably from the nonlinear response of the analog-to-digital converters (ADCNL) of the HFI detectors \cite{Pagano:2019tci}. The \texttt{SRoll2} algorithm \cite{Planck:2016kqe,Pagano:2019tci} provides an improved mapmaking approach that better corrects these systematics, reducing the variance of the low-$\ell$ EE spectrum by a factor of two at $\ell < 6$ compared to the 2018 legacy maps. The resulting \texttt{SRoll2} likelihood has therefore provided tight constraints on $\tau = 0.0566^{+0.0053}_{-0.0062}$ at 68\% confidence level \cite{Pagano:2019tci}, with alternative likelihood methods on the same maps yielding consistent and marginally tighter constraints \cite{deBelsunce:2021mec}.

The \texttt{SRoll2} likelihood is distributed as a pre-tabulated probability table, providing the log-probability of the data as a function of $D_\ell^{EE}$ in units of $10^{-4}\mu{\rm K}^2$ at each multipole from $\ell = 2$ to $29$. As with all other CMB low-$\ell$ likelihoods, these probabilities are non-Gaussian, posing a challenge for analysis methods that assume an analytic Gaussian likelihood. Such methods include the Fisher-bias formalism \cite{Knox:1998fp,Kim:2003mq,Taylor:2006aw,Shapiro:2008yk,DeBernardis:2008tk,Lee:2022gzh} and the Fisher matrix forecasts for future CMB surveys combined with existing \textit{Planck} low-$\ell$ data. In practice, many analyses sidestep this issue by replacing the full \texttt{SRoll2} likelihood with a simple Gaussian prior on $\tau$ or neglecting the low-$\ell$ data altogether. While convenient, this Gaussian prior discards the full shape information of the likelihood including its characteristic asymmetry and the correlation between $\tau$ and $A_s$. Neglecting the low-$\ell$ data, on the other hand, leaves $\tau$ essentially unconstrained, which can affect the inference of correlated parameters such as $A_s$, $H_0$ and the neutrino mass \cite{
Allali:2025wwi,Sailer:2025lxj,Jhaveri:2025neg,Allali:2025yvp}.

The non-Gaussian nature of CMB low-$\ell$ likelihoods and the associated difficulties for parameter estimation have been studied in Ref.~\cite{Bond:1998qg}. The authors proposed the offset log-normal distribution as one of the approximations to the likelihood and showed that it is indeed an accurate approximation near the peak of the likelihood and working in the transformed variable yields an approximately Gaussian $\chi^2$, avoiding the systematic bias that arises from assuming Gaussianity in CMB spectra. This motivated Ref.~\cite{Prince:2021fdv} to perform log-normal fits for the posterior of low-$\ell$ TT and EE spectra compressing the \textit{Planck} \texttt{commander} and \texttt{SimAll} likelihoods to facilitate the use of the likelihood in cosmological analyses.

In this paper, building on these earlier works and extending them in two ways, we present a compressed \emph{Gaussian} likelihood for the \texttt{SRoll2} low-$\ell$ EE data. First, and more importantly, we show that the $\chi^2$ of an offset log-normal likelihood takes a Gaussian form in the log-transformed power spectrum amplitudes and can be used directly in Fisher matrix analyses without any explicit change of variables, as we show in Sec.~\ref{sec:gaussian}. This observation, which to our best knowledge has not been explicitly made before, directly justifies the approach used in earlier work~\cite{Lee:2022gzh,Plombat:2024kla,Mirpoorian:2024fka,Lee:2025yah,Lee:2026hlo} and makes the compressed likelihood directly compatible with Fisher matrix analyses that previously could not incorporate the non-Gaussian low-$\ell$ likelihood. Second, we introduce a piecewise offset log-normal fitting scheme that more accurately captures the asymmetric tails of the binned posteriors, which a single log-normal fails to reproduce. Note that while we focus on low-$\ell$ EE data in this work, the same Gaussian $\chi^2$ structure applies to any offset log-normal compressed likelihood, including the low-$\ell$ TT likelihood of Ref.~\cite{Prince:2021fdv}.

We validate the compressed likelihood against the full \texttt{SRoll2} likelihood by performing MCMC analysis combined with \textit{Planck} and ACT DR6 data, replacing only the low-$\ell$ EE component. We find excellent agreement in the full $\Lambda$CDM posterior across all parameters, with no visible bias in the means and negligible differences in the $1\sigma$ uncertainties. We further demonstrate that Fisher matrix uncertainty estimates on $\tau$ and $\ln(10^{10}A_s)$ from our compressed likelihood agree well with the full MCMC posteriors, and validate the compressed likelihood in extended cosmological models including a non-flat $\Lambda$CDM model and a dark energy model with a time-varying equation of state $(w_0, w_a)$, finding good agreement with the full \texttt{SRoll2} likelihood across all parameters in both cases. Our compressed \emph{Gaussian} likelihood, \texttt{planck-gaussian-lowl}, is available at \href{https://github.com/nanoomlee/planck-gaussian-lowl}{github.com/nanoomlee/planck-gaussian-lowl}. Our package also incorporates the compressed low-$\ell$ TT likelihood of Ref.~\cite{Prince:2021fdv} in a consistent Gaussian $\chi^2$ form.

The paper is organized as follows. In Sec.~\ref{sec:SRoll2} we briefly describe the \texttt{SRoll2} likelihood. In Sec.~\ref{sec:gaussian} we derive the Gaussian $\chi^2$ form of the offset log-normal likelihood and its connection to Fisher matrix analyses. In Sec.~\ref{sec:method} we describe our compression method in detail. In Sec.~\ref{sec:validation} we present the validation of our compressed likelihood. We conclude in Sec.~\ref{sec:conclusion}.

\section{The \texttt{SRoll2} likelihood\\for Planck low-$\ell$ EE data}
\label{sec:SRoll2}

While the compression method we develop is general and applicable to any low-$\ell$ CMB likelihood, we apply it to the \texttt{SRoll2} likelihood, which we briefly describe here. The \texttt{SRoll2} likelihood~\cite{Pagano:2019tci} is based on the High Frequency Instrument (HFI) data of the \textit{Planck} satellite, processed with the improved \texttt{SRoll2} mapmaking algorithm~\cite{Planck:2016kqe}. Compared to the \textit{Planck} 2018 legacy maps, the \texttt{SRoll2} maps contain significantly lower levels of residual large-scale contamination, and provide roughly a factor of two tighter constraint on the reionization optical depth, $\tau = 0.0566^{+0.0053}_{-0.0062}$. In combination with \textit{Planck} and ACT DR6 high-$\ell$ temperature and polarization data, it yields $\tau = 0.0603^{+0.0055}_{-0.0065}$~\cite{AtacamaCosmologyTelescope:2025blo}.

In the publicly available \texttt{SRoll2} likelihood, the posteriors of $D_\ell^{EE} \equiv \ell(\ell+1)C_\ell^{EE}/(2\pi)$ for $2 \leq \ell \leq 29$ are pre-tabulated on a discrete integer grid, $ D_\ell^{EE} \in [0,2999]$ with $D_\ell^{EE}$ in units of $10^{-4}\mu{\rm K}^2$, and used to evaluate the total likelihood given theoretical predictions for the CMB EE power spectrum. While each individual posterior is non-analytic, it roughly follows a log-normal form, motivating the compression approach we describe below.

\section{Gaussian $\chi^2$\\from a log-normal likelihood}
\label{sec:gaussian}

In Bayesian statistics, the posterior distribution of parameters $\theta$ given data $d$ is proportional to the likelihood function and the assumed prior,
\be
p(\theta|d) \propto \mathcal{L}(d|\theta)\, p(\theta).
\ee
For the purpose of this section, $\theta \equiv D_\ell$ denotes the low-$\ell$ CMB angular power spectrum, and we consider the case where the posterior $p(\theta|d)$, obtained with a flat prior on $\theta$, follows an offset log-normal distribution.\footnote{The following arguments hold for the standard log-normal distribution (e.g., when $\theta_0=0$).} We assume the likelihood function takes the offset log-normal form
\be
\mathcal{L}(d|\theta) = \frac{1}{(\theta-\theta_0)\sigma\sqrt{2\pi}} \exp \left[-\frac{(\ln(\theta-\theta_0) - \mu)^2}{2\sigma^2}\right],
\ee
where $\theta_0$ is an offset, and $\mu$ and $\sigma^2$ are the mean and variance of the normal variable $\ln(\theta - \theta_0)$, i.e. $\ln(\theta-\theta_0) \sim \mathcal{N}(\mu, \sigma^2)$. Rewriting this likelihood explicitly,
\begin{eqnarray}
\mathcal{L}(d|\theta) &=& \frac{1}{\sigma\sqrt{2\pi}}\, e^{-\ln(\theta-\theta_0)}\, e^{-[\ln(\theta-\theta_0)-\mu]^2/2\sigma^2} \nonumber\\
&=& \frac{1}{\sigma\sqrt{2\pi}}\, e^{-[\ln(\theta-\theta_0) - \widetilde{\mu}]^2/2\sigma^2 - \sigma^2/2-\widetilde{\mu}},
\end{eqnarray}
where $\widetilde{\mu} \equiv \mu - \sigma^2$. The corresponding $\chi^2$ is
\be
\chi^2(\theta) \equiv -2\ln\mathcal{L}(d|\theta) = \frac{\left[\ln(\theta-\theta_0) - \widetilde{\mu}\right]^2}{\sigma^2},
\label{eq:chi2_theta}
\ee
up to a constant. This is a standard quadratic form in $x \equiv \ln(\theta-\theta_0)$, and its second derivatives with respect to $x$ are those of a Gaussian, since the constant shift $\mu \to \widetilde{\mu}$ drops out exactly under differentiation. Since the Fisher information matrix depends only on second derivatives of the $\chi^2$, it follows that one can use $\chi^2(\theta)$ to compute the Fisher matrix, as long as derivatives are taken with respect to $x = \ln(\theta-\theta_0)$. Explicitly, the Fisher information for $x = \ln(\theta-\theta_0)$ is
\begin{eqnarray}
F_{xx} \equiv \frac{1}{2}\frac{\partial^2\,\chi^2(\theta)}{\partial\left[\ln(\theta-\theta_0)\right]^2}.
\end{eqnarray}
The Fisher matrix for cosmological parameters $p_i$ can then be obtained by
\be
F_{ij} = \sum_{\ell,\ell'} \frac{\partial x_\ell}{\partial p_i}\, F_{x_\ell x_{\ell'}}\, \frac{\partial x_{\ell'}}{\partial p_j},
\ee
where $x_\ell\equiv \ln (\theta_\ell-\theta_{0,\ell})$ and the sum runs over all $\ell$ modes (or bins) included in the likelihood. 

This result has an important practical implication. While the \texttt{SRoll2} likelihood is provided as a non-analytic function of $D_\ell^{EE}$, the compressed \emph{Gaussian} likelihood we construct in Sec.~\ref{sec:method} provides $\chi^2\left(D_\ell^{EE}\right)$ in the quadratic form of Eq.~\eqref{eq:chi2_theta}. This $\chi^2$ can therefore be used directly in Fisher matrix analyses without any explicit change of variables.
 
We note that this way of incorporating the compressed low-$\ell$ EE (and TT) likelihood in the context of the Fisher-bias formalism was introduced in Ref.~\cite{Lee:2022gzh} and has been used in Refs.~\cite{Plombat:2024kla,Mirpoorian:2024fka,Lee:2025yah,Lee:2026hlo}, but without explicitly showing that the second derivatives of $\chi^2(\theta)$ with respect to $x = \ln(D_\ell - D_0)$ are those of a Gaussian, which justifies treating $\ln(D_\ell - D_0)$ as the effective parameter. The result above provides this justification.

In the following section, we show that the \texttt{SRoll2} low-$\ell$ EE likelihood can be well approximated in this quadratic form through piecewise offset log-normal fits to the binned posteriors, accurately capturing the full shape of the likelihood.

\section{Constructing the compressed Gaussian likelihood}
\label{sec:method}

We describe here the construction of our compressed \emph{Gaussian} likelihood for the \textit{Planck} low-$\ell$ EE data. The key observation is that offset log-normal fits to the binned $D_\ell^{EE}$ posteriors yield a $\chi^2$ that takes a Gaussian form in the log-transformed power spectrum amplitudes, and can therefore be used directly in Fisher matrix analyses without any explicit change of variables. We therefore refer to the resulting compressed representation as the \textit{compressed Gaussian likelihood}.

\subsection{Binning scheme}
\label{subsec:binning}

The \texttt{SRoll2} likelihood provides the conditional posterior $p\left(D_\ell^{EE}|D_\ell^{EE, \rm{obs}}\right)$ for each multipole $\ell = 2, \ldots, 29$ separately, with $D_\ell^{EE}$ given in units of $10^{-4}\mu{\rm K}^2$. Hereafter, we simply denote it as $p\left(D_\ell^{EE}\right)$. Rather than working with all 28 individual posteriors directly, we group the multipoles into six bins $\mathcal{B}$,
\be
\mathcal{B} = \bigl\{\{2,3\},\,\{4,5\},\,\{6,7\},\,\{8\text{--}11\},\,\{12\text{--}15\},\,\{16\text{--}29\}\bigr\},
\label{eq:bins}
\ee
 and represent the joint likelihood with a fitted piecewise log-normal function per bin.

For a given bin $i$ containing multipoles $\ell \in \mathcal{B}_i$, the binned log-likelihood is
\be
\ln \mathcal{L}_i \left(\ols{D}^{EE}_i\right) = \sum_{\ell \in \mathcal{B}_i} \ln p \left(D_\ell^{EE} = \ols{D}^{EE}_i\right),
\ee
where $\ols{D}^{EE}_i$ is the mean theoretical EE spectrum $D_\ell^{EE,\,\rm th}$ over the bin,
\be
\ols{D}^{EE}_i \equiv \frac{1}{|\mathcal{B}_i|} \sum_{\ell \in \mathcal{B}_i} D_\ell^{EE,\,\rm th}.
\ee

\begin{figure}[ht!]
\centering
\includegraphics[width=\linewidth,trim=00 10 00 0]{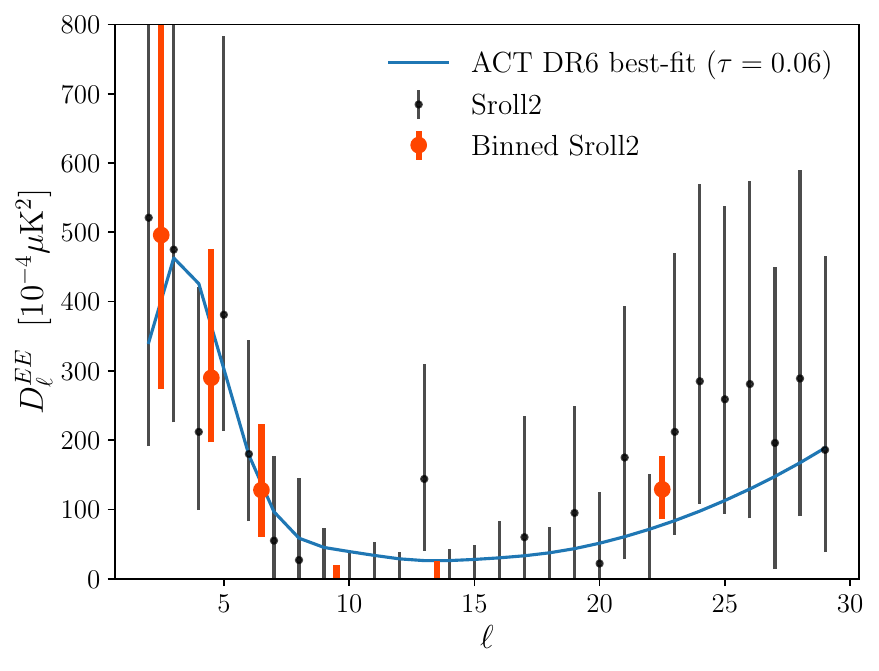}
\caption{The \textit{Planck} low-$\ell$ EE power spectrum from the \texttt{SRoll2} likelihood (black points), compared to the ACT DR6 best-fit $\Lambda$CDM spectrum with $\tau = 0.06$ (blue curve). The black points mark the maximum likelihood values of the individual $D_\ell^{EE}$ posteriors, with error bars indicating the $1\sigma$ credible intervals. The red points show the corresponding quantities for the binned posteriors used in our compressed likelihood. The reionization bump is clearly visible at $\ell \lesssim 6$.}
\vspace{-0.1in}
\label{fig:Dell_EE}
\end{figure}

The choice of binning scheme directly affects the fidelity of the compressed likelihood. For example, while wider bins produce smoother posteriors that are easier to fit, they mix multipoles that may prefer different $D_\ell^{EE}$ values, causing peak shifts and degrading compression accuracy. After exploring many possible binning schemes, we find that the 6-bin scheme defined in Eq.~\eqref{eq:bins} works well in practice. The bins are illustrated in Fig.~\ref{fig:Dell_EE}. We verified that finer binning does not significantly improve the fidelity of the compressed likelihood, mainly because narrower bins produce noisier posteriors that are harder to fit accurately with a log-normal function, resulting in a larger number of parameters without a commensurate gain in accuracy.

We demonstrate in Sec.~\ref{sec:validation} that the 6-bin scheme adopted here accurately reproduces the posterior distributions of all $\Lambda$CDM parameters when combined with \textit{Planck} and ACT DR6 data.

\subsection{Piecewise offset log-normal fit}
\label{subsec:piecewise}

For each bin $i$, we fit the binned log-likelihood $\ln\mathcal{L}_i(\ols{D}^{EE}_i)$ with an offset log-normal function. The offset log-normal likelihood for bin $i$ is
\be
\mathcal{L}_i^{\rm fit} \left(\ols{D}^{EE}_i\right) = \frac{1}{\left(\ols{D}^{EE}_i-D_{0,i}\right)\,\sigma_i\sqrt{2\pi}}\, e^{-\frac{\left[\ln\left(\ols{D}^{EE}_i-D_{0,i}\right)-\mu_i\right]^2}{2\sigma_i^2}},
\label{eq:Lfit}
\ee
where $D_{0,i}$ is an offset ensuring $\ols{D}^{EE}_i > D_{0,i}$, and $\mu_i$ and $\sigma_i$ are parameters characterising the shape of the distribution. Up to a constant, the log of this fitting function takes the form
\be
\ln\mathcal{L}_i^{\rm fit} \left(\ols{D}^{EE}_i\right) = -\frac{\left[\ln\left(\ols{D}^{EE}_i - D_{0,i}\right) - \widetilde{\mu}_i\right]^2}{2\sigma_i^2},
\label{eq:log-Lfit}
\ee
where $\widetilde{\mu}_i \equiv \mu_i - \sigma_i^2$. The offset log-normal provides a natural fitting function for the binned $D_\ell^{EE}$ posteriors because the CMB power spectrum estimators $\hat{C}_\ell$ follow a scaled chi-squared distribution with $2\ell+1$ degrees of freedom, and the log-normal distribution is known to provide an accurate approximation to the chi-squared distribution \cite{Bond:1998qg}.\footnote{Other functional forms such as the offset chi-squared distribution could also serve as fitting functions. However, we use the offset log-normal as it yields a Gaussian $\chi^2$ in $\ln(\overline{D}^{EE}_i - D_{0,i})$ up to a constant, as shown in Sec.~\ref{sec:gaussian} and Eq.~\eqref{eq:log-Lfit}.}

\begin{figure*}[ht!]
\centering
\includegraphics[width = 0.95\linewidth,trim= 00 10 00 10]{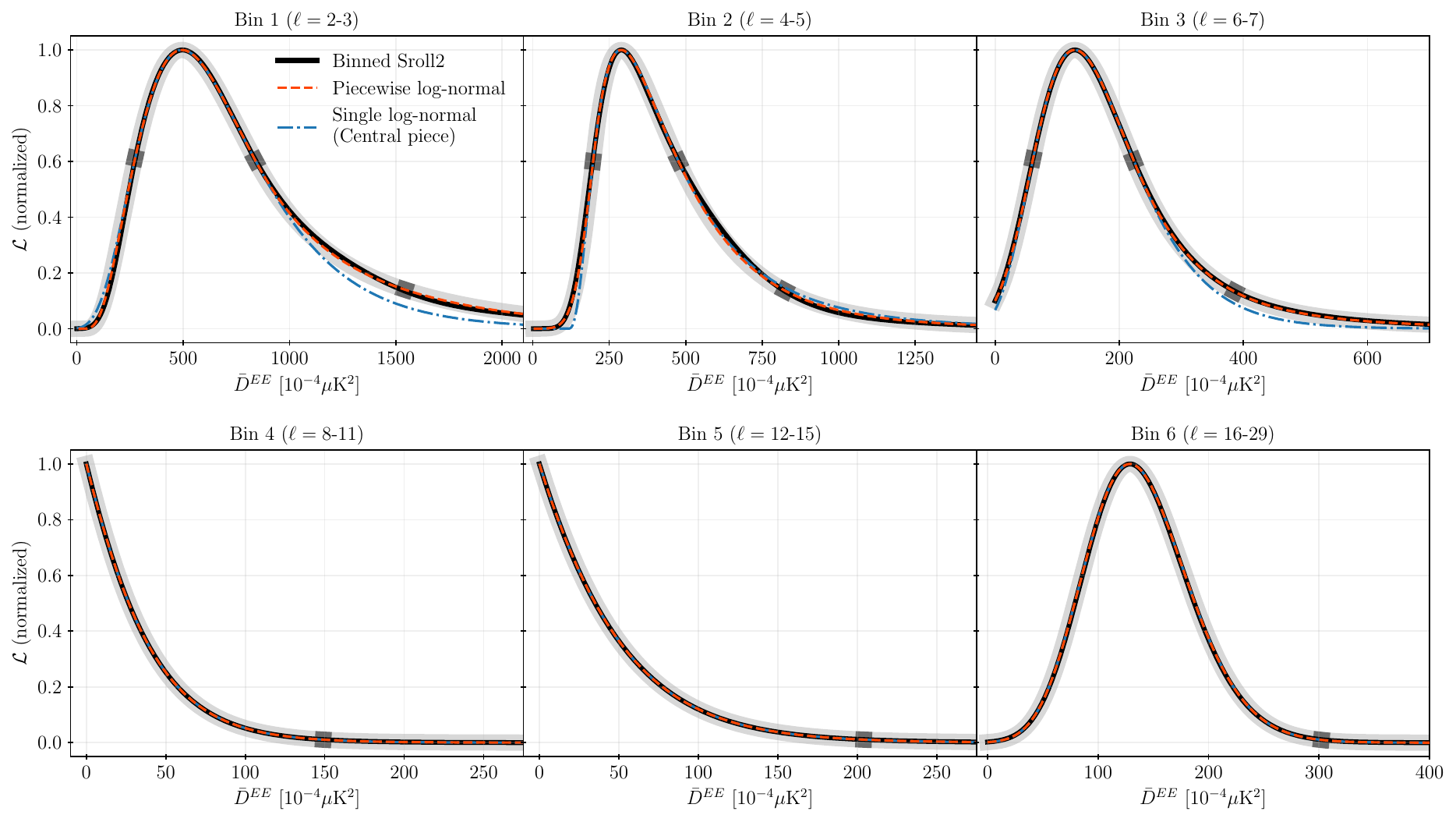}
\caption{Normalized likelihood $\mathcal{L}_i\left(\ols{D}^{EE}_i\right)$ for each of the six bins (black solid), compared to the piecewise offset log-normal fit (orange dashed) and the single offset log-normal fitted over the $\pm1\sigma$ region for bins 1--3 and the $\pm3\sigma$ region for bins 4--6 (blue dash-dotted). The single log-normal fit shown corresponds to the central piece of the piecewise fit. The junction locations between adjacent pieces are indicated by dark grey squares. The piecewise fit accurately captures both the peak and the tails of the true binned posterior, while the single log-normal significantly under(over)estimates the high-$\ols{D}^{EE}$ tail for bins 1 and 3 (bin 2).}
\vspace{-0.1in}
\label{fig:logL}
\end{figure*}

We determine the fit parameters by minimising the likelihood-weighted sum of squared log-likelihood residuals,
\be
\int d\ols{D}^{EE}_i\;\mathcal{L}_i\left(\ols{D}^{EE}_i\right)\left[\ln\mathcal{L}_i\left(\ols{D}^{EE}_i\right) - \ln\mathcal{L}_i^{\rm fit}(\ols{D}^{EE}_i)\right]^2,
\label{eq:loss}
\ee
where the integral is taken over a specified fitting range. This loss function emphasises accuracy where $\mathcal{L}_i$ is large. While a single offset log-normal provides a reasonable approximation near the peak, it fails to capture the correct shape of the tails. This can be seen from the blue dash-dotted curve in the top panels of Fig.~\ref{fig:logL}, which is fitted over the $\pm1\sigma$ region of the binned posterior. 

To address this, we introduce a \textit{piecewise} offset log-normal fit. For each bin, we divide the range of $\ols{D}^{EE}$ into multiple regions, each fitted with a separate offset log-normal of the form Eq.~\eqref{eq:log-Lfit}. One piece containing the peak --- which we call the central piece --- is fitted freely by minimising Eq.~\eqref{eq:loss} over a specified range. All remaining pieces are constrained by requiring that both $\ln\mathcal{L}_i^{\rm fit}$ and its first derivative with respect to $\ols{D}^{EE}$ are continuous at each junction. These two continuity conditions constrain two of the three parameters of each adjacent piece analytically in terms of the parameters of the preceding piece and the junction location, leaving one free parameter to be determined by minimising Eq.~\eqref{eq:loss} over the corresponding region. That is, for an $N$-piece scheme, there are $N+2$ fitting parameters in total.

We adopt the following piecewise scheme.\footnote{Note that the fitting ranges specified here are used only to determine the free parameter of each piece. Once all parameters are determined, the piecewise function is evaluated using the junction locations as boundaries, with the leftmost piece covering all $\overline{D}$ below the left junction and the rightmost piece covering all $\overline{D}$ above the right junction. In cases where a junction crossing falls outside the tabulated \texttt{SRoll2} grid, the fitting range is truncated at the grid boundary.} For bins 1, 2, and 3 ($\ell = 2$--$3$, $4$--$5$, $6$--$7$), which have broader and more asymmetric posteriors, we use a 4-piece scheme where the central piece is fitted over the $(-1\sigma, +1\sigma)$ region:
\begin{itemize}
    \item Piece 1: $(D_{-3\sigma}, D_{-1\sigma})$ --- left tail, 
    constrained from central at $D_{-1\sigma}$
    \item Piece 2: $(D_{-1\sigma}, D_{+1\sigma})$ --- central, freely fitted
    \item Piece 3: $(D_{+1\sigma}, D_{+2\sigma})$ --- constrained from 
    central at $D_{+1\sigma}$
    \item Piece 4: $(D_{+2\sigma}, D_{+5\sigma})$ --- constrained from 
    piece 3 at $D_{+2\sigma}$
\end{itemize}
For bins 4, 5, and 6, which have smoother and more symmetric 
posteriors, we use a 2-piece scheme where the central piece is fitted 
over the $(-3\sigma, +3\sigma)$ region:
\begin{itemize}
    \item Inner: $(D_{-3\sigma}, D_{+3\sigma})$ --- central, freely fitted
    \item Outer: $(D_{+3\sigma}, D_{+5\sigma})$ --- constrained from 
    inner at $D_{+3\sigma}$
\end{itemize}
For the outermost pieces, the fitting range extends to the $5\sigma$ threshold of the binned posterior. For bins 1--3, the posterior does not reach this threshold within the tabulated \texttt{SRoll2} grid, so the fitting range extends to the edge of the probability table. The fitted parameters for all bins are summarised in Table~\ref{tab:params}, and the fitted piecewise offset log-normal fits are shown in Fig.~\ref{fig:logL}. As can be seen in Fig.~\ref{fig:logL}, the piecewise fits accurately capture both the peak and the tails of the true binned posteriors.

\begin{table}[ht!]
\centering
\begin{tabular}{lclcccc}
\hline\hline
Bin & ~Piece~~ & Fitting range & $D_0$ & ~~~~$\widetilde{\mu}$~~~~ & ~~~$\sigma$~~~ \\
\hline
\multirow{4}{*}{\shortstack[l]{Bin 1 \\ ($\ell=2$--$3$)}}
 & 1 & $(72,\ 274)$    & $7.304$    & $6.246$ & $0.666$ \\
 & 2 & $(274,\ 839)$   & $-132.847$ & $6.441$ & $0.436$ \\
 & 3 & $(839,\ 1542)$  & $414.517$  & $5.047$ & $0.999$ \\
 & 4 & $(1542,\ 2999)$ & $-764.308$ & $6.775$ & $0.488$ \\
\hline
\multirow{4}{*}{\shortstack[l]{Bin 2 \\ ($\ell=4$--$5$)}}
 & 1 & $(86,\ 197)$    & $-526.882$ & $6.672$ & $0.087$ \\
 & 2 & $(197,\ 476)$   & $107.656$  & $5.201$ & $0.701$ \\
 & 3 & $(476,\ 824)$   & $47.510$   & $5.452$ & $0.603$ \\
 & 4 & $(824,\ 2999)$  & $151.311$  & $5.123$ & $0.696$ \\
\hline
\multirow{4}{*}{\shortstack[l]{Bin 3 \\ ($\ell=6$--$7$)}}
 & 1 & $(0,\ 61)$      & $-406.681$ & $6.269$ & $0.122$ \\
 & 2 & $(61,\ 223)$    & $-108.511$ & $5.467$ & $0.335$ \\
 & 3 & $(223,\ 386)$   & $66.333$   & $4.341$ & $0.710$ \\
 & 4 & $(386,\ 2999)$  & $133.968$  & $3.721$ & $0.900$ \\
\hline
\multirow{2}{*}{\shortstack[l]{Bin 4 \\ ($\ell=8$--$11$)}}
 & inner         & $(0,\ 149)$     & $-136.446$ & $4.526$ & $0.352$ \\
 & outer         & $(149,\ 514)$   & $-28.144$  & $3.358$ & $0.567$ \\
\hline
\multirow{2}{*}{\shortstack[l]{Bin 5 \\ ($\ell=12$--$15$)}}
 & inner         & $(0,\ 204)$     & $-228.794$ & $4.967$ & $0.333$ \\
 & outer         & $(204,\ 743)$   & $-43.811$  & $3.586$ & $0.581$ \\
\hline
\multirow{2}{*}{\shortstack[l]{Bin 6 \\ ($\ell=16$--$29$)}}
 & inner         & $(15,\ 302)$    & $-208.849$ & $5.822$ & $0.136$ \\
 & outer         & $(302,\ 479)$   & $41.547$   & $4.751$ & $0.267$ \\
\hline\hline
\end{tabular}
\caption{Fitted parameters of the piecewise offset log-normal likelihood for the 6-bin scheme adopted in this work. $D_{0}$, $\widetilde{\mu}$, and $\sigma$ are the parameters of each piece, as defined in Eq.~\eqref{eq:Lfit}. Note that we work with $D_\ell^{EE}$ in units of $10^{-4}\,\mu{\rm K}^2$, and both the fitting ranges and fitted parameters depend on this choice of unit.}
\label{tab:params}
\end{table}

\subsection{Implementation as a Gaussian $\chi^2$}
\label{subsec:gaussian}

As shown in Sec.~\ref{sec:gaussian}, the $\chi^2$ of the offset log-normal likelihood takes a Gaussian form in $\ln(\ols{D}^{EE}_i - D_{0,i})$. For the piecewise scheme, within each piece the log-likelihood takes the quadratic form of Eq.~\eqref{eq:log-Lfit}, so the total $\chi^2$ is
\be
\chi^2 = \sum_{i=1}^{6} \left[\frac{\left[\ln\left(\ols{D}^{EE}_i - D_{0,i}^{(k)}\right) - \widetilde{\mu}_i^{(k)}\right]^2}{\left(\sigma_i^{(k)}\right)^2}+ \Delta_i^{(k)}\right],
\label{eq:chi2}
\ee
where the superscript $(k)$ denotes the piece selected for bin $i$ based on the value of $\ols{D}^{EE}_i$ relative to the junction points, and $\Delta_i^{(k)}$ is a constant offset determined by requiring continuity of the $\chi^2$ at each junction.

For Fisher matrix analyses, which are performed around the peak of the likelihood, it suffices to use only the central piece parameters. Since the central piece is specifically designed to accurately describe the curvature of the log-likelihood near the peak (see Appendix~\ref{app:gaussian}), up to a constant the $\chi^2$ reduces to
\be
\chi^2_{\rm Fisher} = \sum_{i=1}^{6} \frac{\left[\ln \left(\ols{D}^{EE}_i - D_{0,i}^{\rm central}\right) - \widetilde{\mu}_i^{\rm central}\right]^2}{\left(\sigma_i^{\rm central}\right)^2},
\label{eq:chi2_fisher}
\ee
where the superscript ``central'' denotes the parameters of the central piece. This form is directly usable in any analysis framework that assumes a Gaussian $\chi^2$, including the Fisher-bias formalism~\cite{Knox:1998fp,Kim:2003mq,Taylor:2006aw,Shapiro:2008yk,DeBernardis:2008tk,Lee:2022gzh}, by treating $\ln(\ols{D}^{EE}_i - D_{0,i}^{\rm central})$ as the effective parameter for each bin.

\begin{figure*}[ht!]
\centering
\includegraphics[width = 0.95\linewidth,trim= 00 10 00 10]{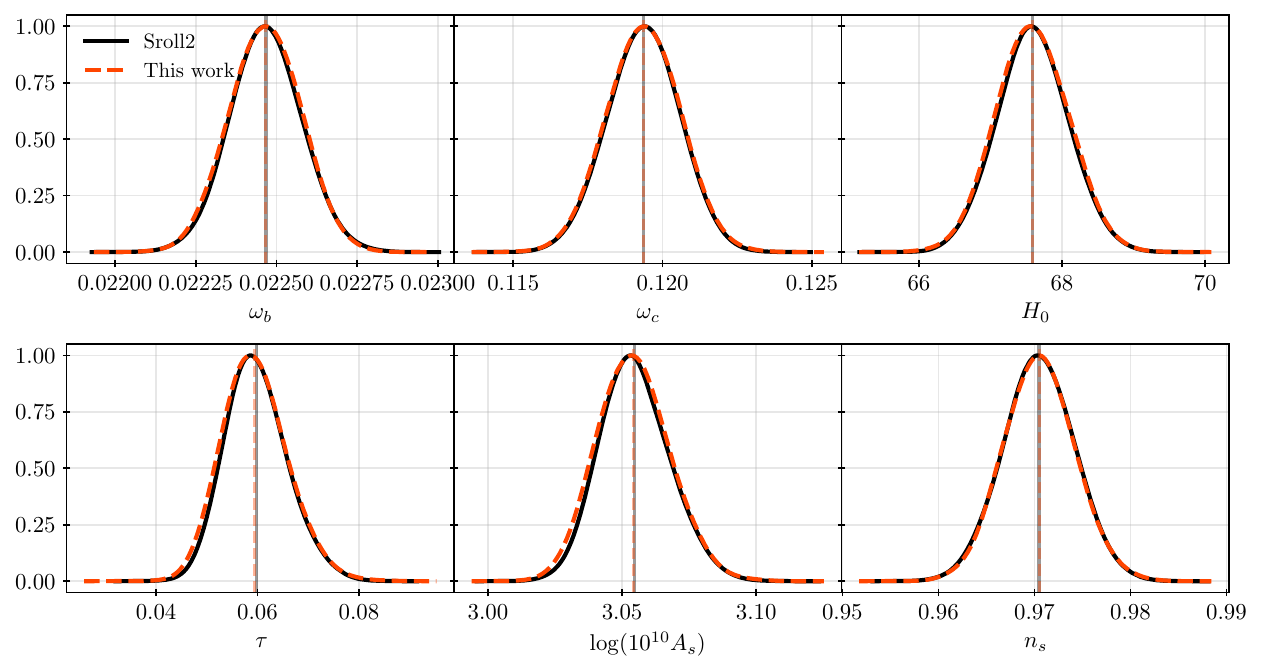}
\caption{Marginalized posterior distributions of the six $\Lambda$CDM parameters from MCMC analyses using the full \texttt{SRoll2} likelihood (black solid) and our compressed \emph{Gaussian} likelihood (orange dashed), combined with \textit{Planck} low-$\ell$ TT, \textit{Planck} high-$\ell$, and ACT DR6 data. Vertical lines indicate the respective posterior means. The two likelihoods are in excellent agreement across all parameters, validating the fidelity of our compression.}
\label{fig:LCDM-posterior}
\end{figure*}

\section{Validation}
\label{sec:validation}

We validate our compressed \emph{Gaussian} likelihood against the \texttt{SRoll2} likelihood by performing MCMC analysis using Cobaya~\cite{Torrado:2020dgo} and \texttt{CLASS}~\cite{class}, assuming the standard flat $\Lambda$CDM model. We use the parameter settings for \texttt{CLASS} suggested by the ACT Collaboration\footnote{\href{https://github.com/ACTCollaboration/DR6_Notebooks/blob/main/ACT_DR6_ps_likelihood.ipynb}{github.com/ACTCollaboration/DR6\_Notebooks/blob/main/ACT\allowbreak\_DR6\_ps\_likelihood.ipynb}} using \texttt{HYREC-2}~\cite{Ali-Haimoud:2010tlj, Ali-Haimoud:2010hou, Lee:2020obi} as the recombination module. In both MCMC runs, we include the \textit{Planck} low-$\ell$ TT likelihood, the \textit{Planck} high-$\ell$ temperature and polarization likelihood, and the ACT DR6 likelihood (ACT-lite), replacing only the low-$\ell$ EE component with either the \texttt{SRoll2} likelihood or our compressed Gaussian likelihood.\footnote{This likelihood combination is equivalent to P-ACT in Ref.~\cite{AtacamaCosmologyTelescope:2025blo}, except that we use ACT-lite instead of the full multi-frequency ACT DR6 likelihood.} We sample all six standard $\Lambda$CDM parameters $\{\omega_b, \omega_c, H_0, \tau, \ln(10^{10}A_s), n_s\}$, along with the ACT and \textit{Planck} calibration parameters $A_{\rm ACT}$ and $A_{\rm Planck}$.\footnote{The \texttt{SRoll2} likelihood rescales the theoretical spectrum at each $\ell$ by dividing by $A_{\rm planck}^2$, where $A_{\rm planck}$ is an overall calibration nuisance parameter. Our compressed likelihood retains this convention for consistency when used in MCMC analyses, with $A_{\rm planck}=1$ as a default value.}

The resulting marginalized posterior distributions are shown in Fig.~\ref{fig:LCDM-posterior}. The posteriors from our compressed likelihood (orange dashed) are in excellent agreement with those from the original \texttt{SRoll2} likelihood (black solid) across all parameters. In particular, $\tau$ and $\ln(10^{10}A_s)$, which are most sensitive to the low-$\ell$ EE likelihood, show no statistically significant bias in the means and negligible differences in the widths. The remaining parameters $\omega_b$, $\omega_c$, $H_0$, and $n_s$, which are primarily constrained by the high-$\ell$ data but are also correlated with $\tau$ and $A_s$, are equally well reproduced. The 2D posteriors are also in excellent agreement, indicating that the compressed likelihood accurately recovers the full parameter covariance from the \texttt{SRoll2} likelihood.

In addition, we demonstrate that our compressed likelihood provides accurate parameter uncertainty estimates through Fisher matrix analyses. Table~\ref{tab:1sigma} presents the $1\sigma$ uncertainties on $\tau$ and $\ln(10^{10}A_s)$ from both Fisher matrix and MCMC, under the same likelihood combination described above. This demonstrates that our compressed Gaussian $\chi^2$, evaluated at the fiducial cosmology, provides Fisher matrix uncertainty estimates in good agreement with the full MCMC posteriors, validating the Gaussian $\chi^2$ formulation of Sec.~\ref{sec:gaussian}. In contrast, computing the Fisher matrix from the \texttt{SRoll2} likelihood assuming Gaussianity in $D_\ell^{EE}$ overestimates the uncertainties, since this assumption does not correctly capture the curvature of the non-Gaussian likelihood in parameter space as expected.

\begin{table}[ht!]
\centering
\begin{tabular}{c|c|c|c}
\hline
& MCMC & $\sigma_{\rm Fisher}$ (this work) & $\sigma_{\rm Fisher}$ (SRoll2) \\
\hline\hline
$\tau$ & $0.0601^{+0.00534}_{-0.00697}$ & $\pm 0.00587$ & $\pm 0.0101$ \\
$\ln(10^{10}A_s)$ & $3.055^{+0.0121}_{-0.0142}$ & $\pm 0.0114$ & $\pm 0.0192$ \\
\hline
\end{tabular}
\caption{Comparison of $1\sigma$ uncertainties on $\tau$ and   $\ln(10^{10}A_s)$ from MCMC and the Fisher matrix, using the same likelihood combination as in Fig.~\ref{fig:LCDM-posterior}. The Fisher matrix uncertainties from our compressed Gaussian likelihood agree well with the MCMC results, while those computed from the \texttt{SRoll2} likelihood assuming Gaussianity in $D_\ell^{EE}$ overestimate the uncertainties. MCMC constraints are quoted as posterior means with asymmetric $68\%$ credible intervals; Fisher matrix uncertainties are symmetric by construction.}
\label{tab:1sigma}
\end{table}

We further validate the compressed likelihood beyond the standard flat $\Lambda$CDM model. For a non-flat $\Lambda$CDM model and a dark energy model with a time-varying equation of state $(w_0, w_a)$, our compressed likelihood yields posteriors consistent with the \texttt{SRoll2} likelihood across all parameters, with mean shifts $\lesssim 0.1\sigma$. We note that these small shifts may partly reflect the fact that extended models can shift the $\tau$ posterior relative to the flat $\Lambda$CDM case, causing the sampler to explore regions where the log-normal approximation is slightly less accurate than near the peak. Nevertheless, the $\lesssim 0.1\sigma$ level agreement demonstrates that the compressed likelihood remains a reliable tool beyond the standard $\Lambda$CDM model. We also emphasize that the primary purpose of this compressed likelihood is not to perfectly reproduce the \texttt{SRoll2} likelihood throughout the entire parameter space, but rather to provide an accurate Gaussian $\chi^2$ near the peak for Fisher matrix analyses.

\section{Conclusion}
\label{sec:conclusion}

We have presented a compressed \emph{Gaussian} likelihood for the \textit{Planck} low-$\ell$ EE polarization data, constructed from the \texttt{SRoll2} likelihood which currently provides the tightest constraint on the reionization optical depth $\tau$. The key observation is that fitting the binned $D_\ell^{EE}$ posteriors with offset log-normal functions [Eq.~\eqref{eq:Lfit}] yields a $\chi^2$ that takes a Gaussian form in the log-transformed power spectrum amplitudes [Eq.~\eqref{eq:chi2}], and can therefore be used directly in Fisher matrix analyses without any explicit change of variables. This makes our compressed likelihood directly compatible with any analysis framework that requires an analytic Gaussian $\chi^2$, without any approximation beyond the log-normal fit itself. Examples include the Fisher-bias formalism~\cite{Knox:1998fp,Kim:2003mq,Taylor:2006aw,Shapiro:2008yk,DeBernardis:2008tk,Lee:2022gzh} and Fisher matrix forecasts for future CMB surveys combined with existing \textit{Planck} low-$\ell$ data.

To accurately capture the asymmetric tails of the binned posteriors, which a single offset log-normal fails to reproduce, we introduce a piecewise fitting scheme in which the range of the binned spectra is divided into multiple regions each fitted with a separate offset log-normal. The central piece containing the peak is fitted freely, and all adjacent pieces are constrained by requiring continuity of both the log-likelihood and its first derivative at each junction, leaving only one free parameter per piece. We fit each piece by minimising the likelihood-weighted sum of squared log-likelihood residuals [Eq.~\eqref{eq:loss}], which prioritises accuracy near the peak of the distribution.

We validate the compressed likelihood against the \texttt{SRoll2} likelihood in three ways. First, we perform MCMC analysis combined with \textit{Planck} and ACT DR6 data under the standard flat $\Lambda$CDM model, replacing only the low-$\ell$ EE component. The compressed likelihood reproduces the posterior distributions of all six $\Lambda$CDM parameters in excellent agreement with the \texttt{SRoll2} likelihood, with no visible bias in the means and negligible differences in the widths (Fig.~\ref{fig:LCDM-posterior}). Second, we demonstrate that the Fisher matrix uncertainties on $\tau$ and $\ln(10^{10}A_s)$ computed from our compressed Gaussian $\chi^2$ agree well with the MCMC posteriors, confirming that the Gaussian $\chi^2$ formulation correctly captures the information content of the low-$\ell$ EE likelihood for Fisher matrix analyses (Table~\ref{tab:1sigma}). Third, we validate the compressed likelihood beyond the standard flat $\Lambda$CDM model, finding consistent posteriors with mean shifts $\lesssim0.1\sigma$ for all parameters in both a non-flat $\Lambda$CDM model and a dark energy model with a time-varying equation of state $(w_0, w_a)$.

Our compressed likelihood, \texttt{planck-gaussian-lowl}, is publicly available at \href{https://github.com/nanoomlee/planck-gaussian-lowl}{github.com/nanoomlee/planck-gaussian-lowl} as a lightweight Python package.\footnote{The repository includes the compressed likelihood module, an example Cobaya likelihood class for MCMC analyses, and an example script for Fisher matrix computation using the central piece parameters [Eq.~\eqref{eq:chi2_fisher}].} The package incorporates the compressed low-$\ell$ TT likelihood of Ref.~\cite{Prince:2021fdv} in a consistent Gaussian $\chi^2$ form, allowing users to straightforwardly combine both compressed low-$\ell$ TT and EE likelihoods. This package provides a simple and accurate tool for cosmological analyses that require a tractable analytic form of the low-$\ell$ EE likelihood, and we expect it to be broadly useful for CMB analyses.

\section*{acknowledgments}
We thank Laura Herold and Benjamin Wandelt for useful discussions. NL was supported by the Horizon Fellowship from Johns Hopkins University. This work was carried out at the Advanced Research Computing at Hopkins (ARCH) core facility (rockfish.jhu.edu), which is supported by the National Science Foundation (NSF) grant number OAC1920103.

\section*{Data Availability}

The data that support the findings of this article are openly available \cite{planck-gaussian-lowl}.

\onecolumngrid

\vspace{0.1in}
\begin{figure}[ht!]
\centering
\includegraphics[width = 0.95\linewidth,trim= 00 10 00 0]{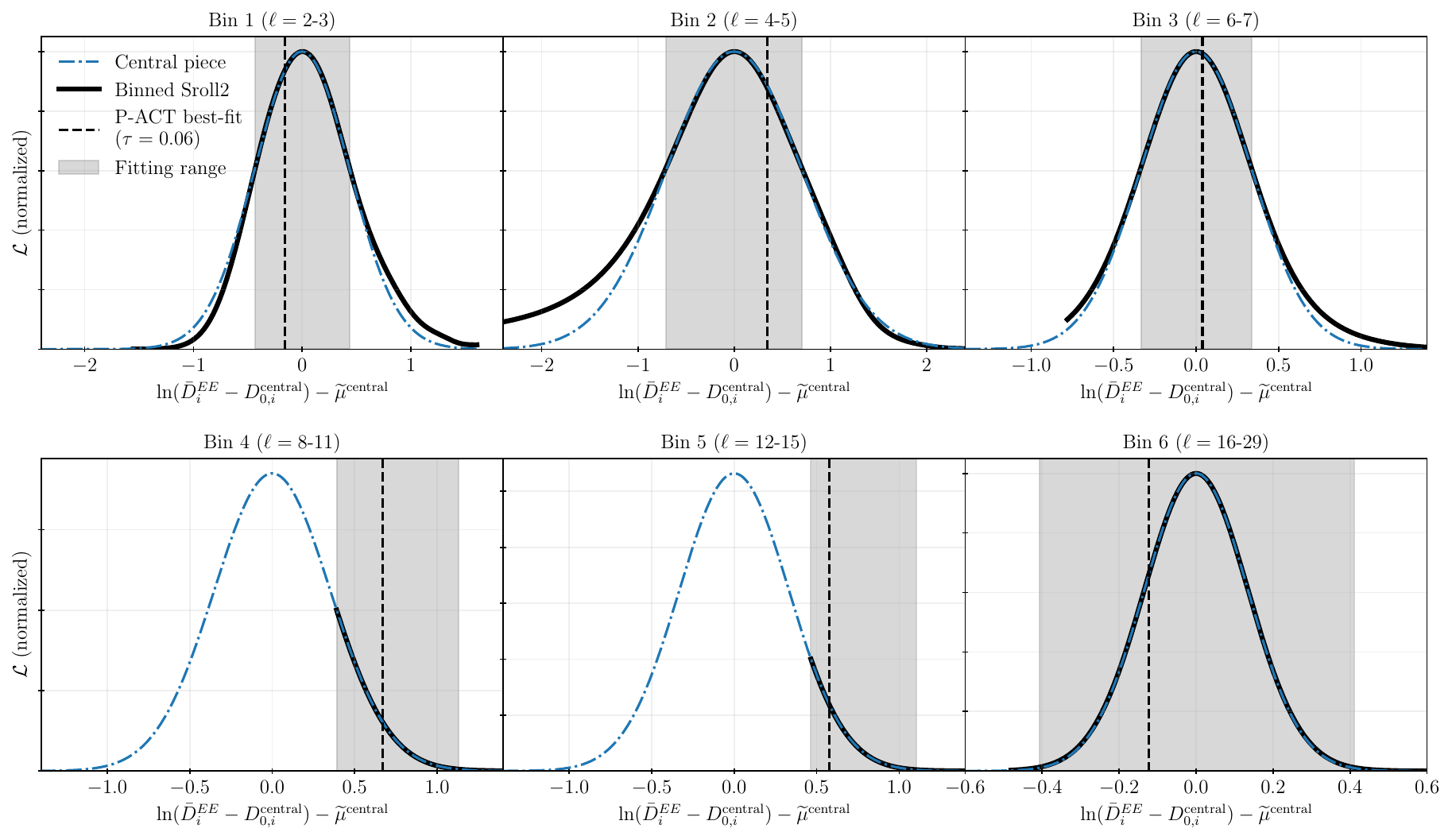}
\caption{Normalized likelihood $\mathcal{L}_i(\overline{D}^{EE}_i)$ for each of the six bins (black solid) as a function of the transformed variable $\ln\left(\ols{D}^{EE}_i - D_{0,i}^{\rm central}\right) - \tilde\mu_i^{\rm central}$, where $D_{0,i}^{\rm central}$ and $\tilde\mu_i^{\rm central}$ are the parameters of the central piece. The central piece (blue dot-dashed) is exactly Gaussian in this variable by construction, peaked at zero, and is extended beyond the fitting range for illustration purposes. The shaded region indicates the fitting range of the central piece, within which the true \texttt{SRoll2} posterior closely follows the Gaussian shape, confirming that the central piece accurately captures the curvature of the likelihood near the peak and validating its use in Fisher matrix analyses [Eq.~\eqref{eq:chi2_fisher}]. The blue vertical line marks the P-ACT best-fit fiducial ($\tau = 0.06$), which falls within the fitting range for all bins.}
\vspace{-0.1in}
\label{fig:L_transf}
\end{figure}

\twocolumngrid

\appendix
\section{Gaussian structure of the central piece}
\label{app:gaussian}

Figure~\ref{fig:L_transf} shows the normalized likelihood $\mathcal{L}_i$ for each bin as a function of the transformed variable $\ln\left(\ols{D}^{EE}_i - D_{0,i}^{\rm central}\right) - \tilde\mu_i^{\rm central}$, where $D_{0,i}^{\rm central}$ and $\tilde\mu_i^{\rm central}$ are the parameters of the central piece. By construction, the central piece (blue, dot-dashed) is exactly Gaussian in this variable, peaked at zero, and is extended beyond the fitting range for illustration purposes. The true \texttt{SRoll2} posterior (black) closely follows this Gaussian shape within the fitting range (shaded region), confirming that the central piece accurately captures the curvature of the likelihood near the peak, validating the use of this central piece to evaluate the Fisher matrix around the peak [Eq.~\eqref{eq:chi2_fisher}]. The blue vertical line shows the P-ACT best-fit fiducial ($\tau = 0.06$), which falls within the fitting range for all bins.

\bibliography{mybib}

@article{Allali:2025yvp,
    author = "Allali, Itamar J. and Li, Lingfeng and Singh, Praniti and Fan, JiJi",
    title = "{Cosmic tensions indirectly correlate with reionization optical depth}",
    eprint = "2509.09678",
    archivePrefix = "arXiv",
    primaryClass = "astro-ph.CO",
    doi = "10.1103/14tf-4nk2",
    journal = "Phys. Rev. D",
    volume = "113",
    number = "8",
    pages = "083515",
    year = "2026"
}

@article{Allali:2025wwi,
    author = "Allali, Itamar J. and Singh, Praniti and Fan, JiJi and Li, Lingfeng",
    title = "{Reionization and the Hubble constant: correlations in the Cosmic Microwave Background}",
    eprint = "2503.05691",
    archivePrefix = "arXiv",
    primaryClass = "astro-ph.CO",
    doi = "10.1088/1475-7516/2025/08/082",
    journal = "JCAP",
    volume = "08",
    pages = "082",
    year = "2025"
}

@misc{planck-gaussian-lowl,
  author       = "Lee, Nanoom",
  title        = "{planck-gaussian-lowl}",
  year         = "2026",
  note         = "Version v1.0, GitHub",
  howpublished = "\url{https://github.com/nanoomlee/planck-gaussian-lowl}"
}

@article{deBelsunce:2021mec,
    author = "de Belsunce, Roger and Gratton, Steven and Coulton, William and Efstathiou, George",
    title = "{Inference of the optical depth to reionization from low multipole temperature and polarization Planck data}",
    eprint = "2103.14378",
    archivePrefix = "arXiv",
    primaryClass = "astro-ph.CO",
    doi = "10.1093/mnras/stab2215",
    journal = "Mon. Not. Roy. Astron. Soc.",
    volume = "507",
    number = "1",
    pages = "1072--1091",
    year = "2021"
}

@article{Bond:1998qg,
    author = "Bond, J. R. and Jaffe, Andrew H. and Knox, L. E.",
    title = "{Radical compression of cosmic microwave background data}",
    eprint = "astro-ph/9808264",
    archivePrefix = "arXiv",
    reportNumber = "CFPA-98-TH-16, CITA-98-25",
    doi = "10.1086/308625",
    journal = "Astrophys. J.",
    volume = "533",
    pages = "19",
    year = "2000"
}

@article{Sailer:2025lxj,
    author = "Sailer, Noah and Farren, Gerrit S. and Ferraro, Simone and White, Martin",
    title = "{Addressing Tensions in {\ensuremath{\Lambda}}CDM Cosmology by an Increase in the Optical Depth to Reionization}",
    eprint = "2504.16932",
    archivePrefix = "arXiv",
    primaryClass = "astro-ph.CO",
    doi = "10.1103/6r54-8lv4",
    journal = "Phys. Rev. Lett.",
    volume = "136",
    number = "8",
    pages = "081002",
    year = "2026"
}

@article{Jhaveri:2025neg,
    author = "Jhaveri, Tanisha and Karwal, Tanvi and Hu, Wayne",
    title = "{Turning a negative neutrino mass into a positive optical depth}",
    eprint = "2504.21813",
    archivePrefix = "arXiv",
    primaryClass = "astro-ph.CO",
    doi = "10.1103/6vd2-rbfn",
    journal = "Phys. Rev. D",
    volume = "112",
    number = "4",
    pages = "043541",
    year = "2025"
}

@article{Plombat:2024kla,
    author = "Plombat, Hugo and Simon, Th{\'e}o and Flitter, Jordan and Poulin, Vivian",
    title = "{Probing dark relativistic species and their interactions with dark matter through CMB and 21 cm surveys}",
    eprint = "2410.01486",
    archivePrefix = "arXiv",
    primaryClass = "astro-ph.CO",
    doi = "10.1088/1475-7516/2025/01/071",
    journal = "JCAP",
    volume = "01",
    pages = "071",
    year = "2025"
}

@article{Planck:2018lkk,
    author = "Aghanim, N. and others",
    collaboration = "Planck",
    title = "{Planck 2018 results. III. High Frequency Instrument data processing and frequency maps}",
    eprint = "1807.06207",
    archivePrefix = "arXiv",
    primaryClass = "astro-ph.CO",
    doi = "10.1051/0004-6361/201832909",
    journal = "Astron. Astrophys.",
    volume = "641",
    pages = "A3",
    year = "2020"
}

@article{Planck:2019nip,
    author = "Aghanim, N. and others",
    collaboration = "Planck",
    title = "{Planck 2018 results. V. CMB power spectra and likelihoods}",
    eprint = "1907.12875",
    archivePrefix = "arXiv",
    primaryClass = "astro-ph.CO",
    doi = "10.1051/0004-6361/201936386",
    journal = "Astron. Astrophys.",
    volume = "641",
    pages = "A5",
    year = "2020"
}

@article{Planck:2018nkj,
    author = "Aghanim, N. and others",
    collaboration = "Planck",
    title = "{Planck 2018 results. I. Overview and the cosmological legacy of Planck}",
    eprint = "1807.06205",
    archivePrefix = "arXiv",
    primaryClass = "astro-ph.CO",
    doi = "10.1051/0004-6361/201833880",
    journal = "Astron. Astrophys.",
    volume = "641",
    pages = "A1",
    year = "2020"
}

@article{Lee:2026hlo,
    author = "Lee, Nanoom and Zhou, Tianji",
    title = "{What it takes to solve the Hubble tension through Modifications of Cosmological Recombination II: in light of ACT DR6 and DESI DR2}",
    eprint = "2606.06495",
    archivePrefix = "arXiv",
    primaryClass = "astro-ph.CO",
    month = "6",
    year = "2026"
}

@article{Planck:2016kqe,
    author = "Aghanim, N. and others",
    collaboration = "Planck",
    title = "{Planck intermediate results. XLVI. Reduction of large-scale systematic effects in HFI polarization maps and estimation of the reionization optical depth}",
    eprint = "1605.02985",
    archivePrefix = "arXiv",
    primaryClass = "astro-ph.CO",
    doi = "10.1051/0004-6361/201628890",
    journal = "Astron. Astrophys.",
    volume = "596",
    pages = "A107",
    year = "2016"
}

@article{Pagano:2019tci,
    author = "Pagano, L. and Delouis, J. -M. and Mottet, S. and Puget, J. -L. and Vibert, L.",
    title = "{Reionization optical depth determination from Planck HFI data with ten percent accuracy}",
    eprint = "1908.09856",
    archivePrefix = "arXiv",
    primaryClass = "astro-ph.CO",
    doi = "10.1051/0004-6361/201936630",
    journal = "Astron. Astrophys.",
    volume = "635",
    pages = "A99",
    year = "2020"
}

@article{AtacamaCosmologyTelescope:2025blo,
    author = "Louis, Thibaut and others",
    collaboration = "Atacama Cosmology Telescope",
    title = "{The Atacama Cosmology Telescope: DR6 power spectra, likelihoods and {\ensuremath{\Lambda}}CDM parameters}",
    eprint = "2503.14452",
    archivePrefix = "arXiv",
    primaryClass = "astro-ph.CO",
    reportNumber = "FERMILAB-PUB-25-0071-PPD",
    doi = "10.1088/1475-7516/2025/11/062",
    journal = "JCAP",
    volume = "11",
    pages = "062",
    year = "2025"
}

@article{Torrado:2020dgo,
    author = "Torrado, Jesus and Lewis, Antony",
    title = "{Cobaya: Code for Bayesian Analysis of hierarchical physical models}",
    eprint = "2005.05290",
    archivePrefix = "arXiv",
    primaryClass = "astro-ph.IM",
    reportNumber = "TTK-20-15",
    doi = "10.1088/1475-7516/2021/05/057",
    journal = "JCAP",
    volume = "05",
    pages = "057",
    year = "2021"
}

@article{Mirpoorian:2024fka,
    author = "Mirpoorian, Seyed Hamidreza and Jedamzik, Karsten and Pogosian, Levon",
    title = "{Modified recombination and the Hubble tension}",
    eprint = "2411.16678",
    archivePrefix = "arXiv",
    primaryClass = "astro-ph.CO",
    month = "11",
    year = "2024"
}

@article{DeBernardis:2008tk,
    author = "De Bernardis, Francesco and Bean, Rachel and Galli, Silvia and Melchiorri, Alessandro and Silk, Joseph I. and Verde, Licia",
    title = "{Delayed Recombination and Standard Rulers}",
    eprint = "0812.3557",
    archivePrefix = "arXiv",
    primaryClass = "astro-ph",
    doi = "10.1103/PhysRevD.79.043503",
    journal = "Phys. Rev. D",
    volume = "79",
    pages = "043503",
    year = "2009"
}

@article{Shapiro:2008yk,
    author = "Shapiro, Charles",
    title = "{Biased Dark Energy Constraints from Neglecting Reduced Shear in Weak Lensing Surveys}",
    eprint = "0812.0769",
    archivePrefix = "arXiv",
    primaryClass = "astro-ph",
    doi = "10.1088/0004-637X/696/1/775",
    journal = "Astrophys. J.",
    volume = "696",
    pages = "775--784",
    year = "2009"
}

@article{Taylor:2006aw,
    author = "Taylor, A. N. and Kitching, Thomas D. and Bacon, D. J. and Heavens, A. F.",
    title = "{Probing dark energy with the shear-ratio geometric test}",
    eprint = "astro-ph/0606416",
    archivePrefix = "arXiv",
    doi = "10.1111/j.1365-2966.2006.11257.x",
    journal = "Mon. Not. Roy. Astron. Soc.",
    volume = "374",
    pages = "1377--1403",
    year = "2007"
}

@article{Kim:2003mq,
    author = "Kim, Alex G. and Linder, Eric V. and Miquel, Ramon and Mostek, Nick",
    title = "{Effects of systematic uncertainties on the supernova determination of cosmologial parameters}",
    eprint = "astro-ph/0304509",
    archivePrefix = "arXiv",
    doi = "10.1111/j.1365-2966.2004.07260.x",
    journal = "Mon. Not. Roy. Astron. Soc.",
    volume = "347",
    pages = "909--920",
    year = "2004"
}

@article{Knox:1998fp,
    author = "Knox, Lloyd and Scoccimarro, Roman and Dodelson, Scott",
    title = "{The Impact of inhomogeneous reionization on cosmic microwave background anisotropy}",
    eprint = "astro-ph/9805012",
    archivePrefix = "arXiv",
    reportNumber = "CITA-98-13, FERMILAB-PUB-98-405-A",
    doi = "10.1103/PhysRevLett.81.2004",
    journal = "Phys. Rev. Lett.",
    volume = "81",
    pages = "2004--2007",
    year = "1998"
}

@article{Prince:2021fdv,
    author = "Prince, Heather and Dunkley, Jo",
    title = "{Compressed Python likelihood for large scale temperature and polarization from Planck}",
    eprint = "2104.05715",
    archivePrefix = "arXiv",
    primaryClass = "astro-ph.CO",
    doi = "10.1103/PhysRevD.105.023518",
    journal = "Phys. Rev. D",
    volume = "105",
    number = "2",
    pages = "023518",
    year = "2022"
}

@article{Ali-Haimoud:2010tlj,
    author = "Ali-Haimoud, Yacine and Hirata, Christopher M.",
    title = "{Ultrafast effective multi-level atom method for primordial hydrogen recombination}",
    eprint = "1006.1355",
    archivePrefix = "arXiv",
    primaryClass = "astro-ph.CO",
    doi = "10.1103/PhysRevD.82.063521",
    journal = "Phys. Rev. D",
    volume = "82",
    pages = "063521",
    year = "2010"
}

@article{Planck2018,
    author = "Aghanim, N. and others",
    collaboration = "Planck",
    title = "{Planck 2018 results. VI. Cosmological parameters}",
    eprint = "1807.06209",
    archivePrefix = "arXiv",
    primaryClass = "astro-ph.CO",
    doi = "10.1051/0004-6361/201833910",
    journal = "Astron. Astrophys.",
    volume = "641",
    pages = "A6",
    year = "2020"
}

@article{Ali-Haimoud:2010hou,
    author = "Ali-Haimoud, Yacine and Hirata, Christopher M.",
    title = "{HyRec: A fast and highly accurate primordial hydrogen and helium recombination code}",
    eprint = "1011.3758",
    archivePrefix = "arXiv",
    primaryClass = "astro-ph.CO",
    doi = "10.1103/PhysRevD.83.043513",
    journal = "Phys. Rev. D",
    volume = "83",
    pages = "043513",
    year = "2011"
}

@article{Lee:2020obi,
    author = {Lee, Nanoom and Ali-Ha\"\i{}moud, Yacine},
    title = "{HYREC-2: a highly accurate sub-millisecond recombination code}",
    eprint = "2007.14114",
    archivePrefix = "arXiv",
    primaryClass = "astro-ph.CO",
    doi = "10.1103/PhysRevD.102.083517",
    journal = "Phys. Rev. D",
    volume = "102",
    number = "8",
    pages = "083517",
    year = "2020"
}

@article{class,
    author = "Blas, Diego and Lesgourgues, Julien and Tram, Thomas",
    title = "{The Cosmic Linear Anisotropy Solving System (CLASS) II: Approximation schemes}",
    eprint = "1104.2933",
    archivePrefix = "arXiv",
    primaryClass = "astro-ph.CO",
    reportNumber = "CERN-PH-TH-2011-082, LAPTH-010-11",
    doi = "10.1088/1475-7516/2011/07/034",
    journal = "JCAP",
    volume = "07",
    pages = "034",
    year = "2011"
}

@article{Lee:2022gzh,
	author = {Lee, Nanoom and Ali-Ha\"\i{}moud, Yacine and Sch\"oneberg, Nils and Poulin, Vivian},
	title = "{What It Takes to Solve the Hubble Tension through Modifications of Cosmological Recombination}",
	eprint = "2212.04494",
	archivePrefix = "arXiv",
	primaryClass = "astro-ph.CO",
	doi = "10.1103/PhysRevLett.130.161003",
	journal = "Phys. Rev. Lett.",
	volume = "130",
	number = "16",
	pages = "161003",
	year = "2023"
}

@article{Lee:2025yah,
    author = {Lee, Nanoom and Braglia, Matteo and Ali-Ha{\"\i}moud, Yacine},
    title = "{What it takes to solve the Hubble tension through scale-dependent modifications of the primordial power spectrum}",
    eprint = "2504.07966",
    archivePrefix = "arXiv",
    primaryClass = "astro-ph.CO",
    doi = "10.1103/9q3f-5zrd",
    journal = "Phys. Rev. D",
    volume = "112",
    number = "8",
    pages = "083506",
    year = "2025"
}

\end{document}